\documentclass[pra]{revtex4-2}
 \usepackage{amssymb} \usepackage{graphicx}
\usepackage[utf8]{inputenc}
\usepackage[english]{babel} 
\usepackage{amsthm}
 \usepackage{amsmath}

\begin{document}
 \title{Symmetries of modified Dirac operators in supergravity flux backgrounds}

\author{\"{U}mit Ertem}
\email{umitertemm@gmail.com}
\address{Diyanet \.{I}\c{s}leri Ba\c{s}kanl{\i}\u{g}{\i}, Astronomy Unit, \"{U}niversiteler Mah.\\
 Dumlup{\i}nar Bul. No:147/H 06800 \c{C}ankaya, Ankara, Turkey}

\begin{abstract}

Modifications of Dirac operators in supergravity flux backgrounds are considered. Modified spin curvature operators and squares of modified Dirac operators corresponding to Schr\"odinger-Lichnerowicz-like formulas are obtained for different types of flux modifications. Symmetry operators of modified massless and massive Dirac equations are found in terms of modified Killing-Yano and modified conformal Killing-Yano forms. Extra constraints for symmetry operators in terms of different types of fluxes and modified Killing-Yano forms are determined.

\end{abstract}

\maketitle

\section{Introduction}

The motion of relativistic particles are described by massless or massive Dirac equation. For general backgrounds, they can be written in terms of the Dirac operator in spin geometry \cite{Lawson Michelsohn, Benn Tucker, Friedrich}. Finding symmetry operators of the Dirac operator can result to find the solutions of the Dirac equation. Symmetry operators take a solution of the equation and give another solution. If the symmetry operators satisfy an algebra structure, then one can obtain the solutions of the equation by separation of variables \cite{Miller, Cariglia, Frolov Kubiznak}. Symmetry operators of the massive Dirac equation can be written in terms of the hidden symmetries of the background which are called Killing-Yano (KY) forms \cite{Benn Kress, Acik Ertem Onder Vercin}. KY forms are antisymmetric generalizations of the Killing vector fields which generate the isometries of the background. Similarly, symmetry operators of the massless Dirac equation can be obtained by using conformal Killing-Yano (CKY) forms which are antisymmetric generalizations of conformal Killing vector fields \cite{Benn Charlton Kress, Benn Charlton}. Besides the non-interacting case, the symmetry operators of the Dirac equation in the presence of electromagnetic interaction are also studied in the literature \cite{Acik Ertem Onder Vercin}. The symmetry operators in the presence of interactions can again be written in terms of the KY forms, however in that case there are extra constraints on KY forms to satisfy the symmetry operator property \cite{Kubiznak Warnick Krtous}.

In supergravity backgrounds which correspond to the solutions of supergravity theories in various dimensions, one needs to define extra flux forms besides metric. In the presence of fluxes, the Dirac equation is modified depending on the type of the flux. This corresponds to the generalization of the modification of the Dirac equation by electromagnetic potential to higher degree fluxes. Some special types of fluxes are considered in the literature for the modification of the Dirac equation and the effect on the symmetry operators \cite{Kubiznak Warnick Krtous}. In the literature, only odd degree fluxes are considered and complicated anomaly equations are found as constraints for symmetry operators. However, these fluxes can be considered in a general way to include all types of supergravity theories in one context. The modification of the Dirac equation by all types of fluxes can be studied via modified Dirac operators \cite{Gutowski Papadopoulos}.

In this paper, we consider the modified Dirac operators that include modifications by various degrees of flux forms. By defining modified spin curvature operators and calculating the Schr\"odinger-Lichnerowicz-like formulas for modified Dirac operators, we investigate the symmetry operators for modified Dirac operators for different modifications. We consider both the massless and massive cases and obtain the symmetry operators of the modified Dirac equation in terms of modified KY and modified CKY forms with some extra constraints. Our analysis of symmetry operators of the modified Dirac equation generalizes the previous unmodified and special cases to all types of modifications in one general context. The constraints to construct symmetry operators are obtained in a simplified and consistent form. Moreover, we also define the modifications of hidden symmetries in the presence of fluxes and obtain the subclasses of modified hidden symmetries that are related to the symmetry operators of the modified Dirac equation.

The paper is organized as follows. In Section II, we define the modified spin connections and modified Dirac operators in the presence of different types of fluxes and we also construct modified spin curvature operators with squares of modified Dirac operators as generalizations of the Schr\"odinger-Lichnerowicz formula. In Section III, we obtain the symmetry operators for modified massless Dirac equation whose solutions correspond to the modified harmonic spinors in terms of the modified CKY forms. Section IV includes the symmetry operators for modified massive Dirac equation in terms of modified KY forms. Section V concludes the paper. In an Appendix, we include the detailed calculations for modified Dirac operators, spin curvature operators and squares of Dirac operators for the special cases.

\section{Modified spin connections and Dirac operators}

Let us consider an $n$-dimensional spin manifold $M$. In terms of the frame basis $\{X_a\}$, the dual coframe basis $\{e^a\}$ and the spin connection $\nabla$ acting on spinor fields, we can define the Dirac operator $\displaystyle{\not}D$ as
\begin{equation}
\displaystyle{\not}D=e^a.\nabla_{X_a}
\end{equation}
where $.$ denotes the Clifford product. The spin connection and the Dirac operator can be generalized to include extra terms written in terms of the various degree Clifford forms. These are called modified spin connections and modified Dirac operators which are encountered in supergravity theories in general. In the presence of a 1-form gauge field $A$ in the background, we have 1-form modified or gauged Dirac operators. For various fluxes present in different supergravity theories, we need to modify the Dirac operator to include these fluxes in the geometry. 2-form flux in five dimensional supergravity, 3-form flux in six or ten dimensional supergravities and 4-form flux in eleven dimensional supergravity will be considered in the paper. We investigate symmetry operators of modified Dirac operators and extend the investigations of \cite{Acik Ertem Onder Vercin, Kubiznak Warnick Krtous} to include all possible cases.  In general, the action of a $p$-form modified spin connection $\widehat{\nabla}$ on a spinor field $\psi$ for a $p$-form $\alpha$ is defined as
\begin{equation}
\widehat{\nabla}_X\psi=\nabla_X\psi+k_1\alpha.\widetilde{X}.\psi+k_2(i_X\alpha).\psi
\end{equation}
where $i_X$ is the contraction operator with respect to the vector field $X$, $k_1$ and $k_2$ are constants and $\widetilde{X}$ denotes the metric dual of the vector field $X$. Then, by considering the definition of the Dirac operator in (1), the $p$-form modified Dirac operator is written as follows
\begin{eqnarray}
\widehat{\displaystyle{\not}D}\psi&=&e^a.\widehat{\nabla}_{X_a}\psi\nonumber\\
&=&\displaystyle{\not}D\psi+k_1e^a.\alpha.e_a.\psi+k_2e^a.i_{X_a}\alpha.\psi\nonumber\\
&=&\displaystyle{\not}D\psi+k_1(-1)^p(n-2p)\alpha.\psi+k_2p\alpha.\psi
\end{eqnarray}
where we have used the Clifford algebra identities $e^a.\alpha.e_a=(-1)^p(n-2p)\alpha$ and $e^a.i_{X_a}\alpha=p\alpha$ for a $p$-form $\alpha$.

For the Levi-Civita connection $\nabla$, the curvature operator is defined with respect to the vector fields $X$ and $Y$ as
\begin{equation}
R(X,Y)=\nabla_X\nabla_Y-\nabla_Y\nabla_X-\nabla_{[X,Y]}
\end{equation}
where $[\,,\,]$ denotes the Lie bracket of vector fields. The same is true for the spin connection and the action of the spin curvature operator on a spinor $\psi$ can be written in terms of the curvature 2-forms $R_{ab}$ as follows \cite{Benn Tucker}
\begin{equation}
R(X_a,X_b)\psi=\frac{1}{2}R_{ab}.\psi.
\end{equation}
From the modified spin connections given above, we can also define the modified spin curvature operator as
\begin{equation}
\widehat{R}(X,Y)=\widehat{\nabla}_X\widehat{\nabla}_Y-\widehat{\nabla}_Y\widehat{\nabla}_X-\widehat{\nabla}_{[X,Y]}
\end{equation}
and by using the definitions of modified spin connections, we can also calculate the action of the modified spin curvature operators on spinor fields. For a general $p$-form modified spin connection defined in (2), we can write the modified spin curvature operator as
\begin{eqnarray}
\widehat{R}(X,Y)\psi&=&\widehat{\nabla}_X(\nabla_Y\psi+k_1\alpha.\widetilde{Y}.\psi+k_2(i_Y\alpha).\psi)\nonumber\\
&&-\widehat{\nabla}_Y(\nabla_X\psi+k_1\alpha.\widetilde{X}.\psi+k_2(i_X\alpha).\psi)\nonumber\\
&&-\nabla_{[X,Y]}-k_1\alpha.\widetilde{[X,Y]}.\psi-k_2(i_{[X,Y]}\alpha).\psi\nonumber\\
&=&R(X,Y)\psi+k_1(\nabla_X\alpha.\widetilde{Y}-\nabla_Y\alpha.\widetilde{X}).\psi+k_1\alpha.(\nabla_X\widetilde{Y}-\nabla_Y\widetilde{X}).\psi\nonumber\\
&&+k_2(\nabla_Xi_Y-\nabla_Yi_X)\alpha.\psi+k_1^2\alpha.(\widetilde{X}.\alpha.\widetilde{Y}-\widetilde{Y}.\alpha.\widetilde{X}).\psi+k_1k_2[i_X\alpha, \alpha.\widetilde{Y}]_{Cl}.\psi\nonumber\\
&&+k_1k_2[\alpha.\widetilde{X}, i_Y\alpha]_{Cl}.\psi+k_2^2[i_X\alpha, i_Y\alpha]_{Cl}.\psi-k_1\alpha.\widetilde{[X,Y]}.\psi-k_2(i_{[X,Y]}\alpha).\psi
\end{eqnarray}
where $[\,,\,]_{Cl}$ denotes the Clifford bracket and in the normal coordinates with the properties $\nabla_{X_a}X_b=0=[X_a, X_b]$, we have
\begin{eqnarray}
\widehat{R}(X_a,X_b)\psi&=&R(X_a,X_b)\psi+k_1(\nabla_{X_a}\alpha.e_b-\nabla_{X_b}\alpha.e_a).\psi+k_1^2\alpha.(e_a.\alpha.e_b-e_b.\alpha.e_a).\psi\nonumber\\
&&+k_1k_2[i_{X_a}\alpha, \alpha.e_b]_{Cl}.\psi+k_2(\nabla_{X_a}i_{X_b}-\nabla_{X_b}i_{X_a})\alpha.\psi+k_1k_2[\alpha.e_a, i_{X_b}\alpha]_{Cl}.\psi\nonumber\\
&&+k_2^2[i_{X_a}\alpha, i_{X_b}\alpha]_{Cl}.\psi.
\end{eqnarray}

By using the definitions of modified Dirac operators and the actions of modified spin curvature operators on spinors, we can calculate the squares of modified Dirac operators which correspond to the Schr\"odinger-Lichnerowicz formulas for the modified cases. For the ordinary Dirac operator, the Schr\"odinger-Lichnerowicz formula reads as
\begin{equation}
\displaystyle{\not}D^2\psi=\nabla^2\psi-\frac{1}{4}\mathcal{R}\psi
\end{equation}
where $\nabla^2$ is the Laplacian acting on spinors and $\mathcal{R}$ is the curvature scalar. By direct computation, we can find the relevant generalizations of the Schr\"odinger-Lichnerowicz formula to the modified cases. For the general $p$-form modified case, we have
\begin{eqnarray}
\widehat{\displaystyle{\not}D}^2\psi&=&\widehat{\nabla}^2\psi+\frac{1}{2}e^{ab}.\widehat{R}(X_a, X_b)\psi\nonumber\\
&=&\widehat{\nabla}^2\psi+\frac{1}{4}e^{ab}.R_{ab}.\psi+\frac{k_1}{2}e^{ab}.(\nabla_{X_a}\alpha.e_b-\nabla_{X_b}\alpha.e_a).\psi+\frac{k_2}{2}e^{ab}.(\nabla_{X_a}i_{X_b}\alpha-\nabla_{X_b}i_{X_a}\alpha).\psi\nonumber\\
&&+\frac{k_1^2}{2}e^{ab}.\alpha.(e_a.\alpha.e_b-e_b.\alpha.e_a).\psi+\frac{k_1k_2}{2}e^{ab}.([\alpha.e_a,i_{X_b}\alpha]_{Cl}+[i_{X_a}\alpha,\alpha.e_b]_{Cl}).\psi\nonumber\\
&&+\frac{k_2^2}{2}e^{ab}.[i_{X_a}\alpha,i_{X_b}\alpha]_{Cl}.\psi.
\end{eqnarray}
For the modified Dirac operator, spin curvature operator and square of the Dirac operator in (3), (8) and (10), special cases of $p=0,1,2,3,4$ will be used in the paper. Explicit calculations of those operators for these special cases are given in the Appendix.

\section{Symmetry operators for modified harmonic spinors}

Let us consider a $\mathbb{Z}_2$-homogeneous first-order differential operator
\begin{equation}
L=\omega^a.\widehat{\nabla}_{X_a}+\Omega
\end{equation}
written in terms of the modified spin connection $\widehat{\nabla}$ and polyforms (inhomogeneous forms) $\omega^a$ and $\Omega$. Because of the $\mathbb{Z}_2$-homogeneous nature of $L$, the polyforms $\omega^a$ and $\Omega$ are both even or both odd differential forms which can contain multiple even (or odd) forms. We will consider modified harmonic spinors $\psi$ which satisfy the equation
\begin{equation}
\widehat{\displaystyle{\not}D}\psi=0.
\end{equation}
If the operator $L$ satisfies the following condition
\begin{equation}
[\widehat{\displaystyle{\not}D}, L]_{gCl} = M\widehat{\displaystyle{\not}D}
\end{equation}
for any operator
\begin{equation}
M=2m^a.\widehat{\nabla}_{X_a}+m
\end{equation}
written in terms of differential forms $m^a$ and $m$, then $L$ defines a symmetry operator for modified harmonic spinors. This means that if $\psi$ is a modified harmonic spinor, then $L\psi$ is also a modified harmonic spinor. Here, $[\,,\,]_{gCl}$ is the graded Clifford bracket defined for a 1-form $\alpha$ and an inhomogeneous Clifford form $\beta$ as \cite{Acik Ertem Onder Vercin}
\begin{equation}
[\alpha, \beta]_{gCl}=\alpha.\beta - \beta^{\eta}.\alpha
\end{equation}
where $\eta$ is the main involution of the Clifford algebra and it acts on a $p$-form $\omega$ as $\omega^{\eta}=(-1)^p\omega$.

Now, we can calculate the left and right hand sides of (13) for the operator (11) and modified Dirac operators. By substituting (11) in (13), the left hand side of (13) acting on a spinor $\psi$ will correspond to
\begin{eqnarray}
[\widehat{\displaystyle{\not}D}, L]_{gCl}\psi&=&\widehat{\displaystyle{\not}D}L\psi-L^{\eta}\widehat{\displaystyle{\not}D}\psi\nonumber\\
&=&e^a.\widehat{\nabla}_{X_a}(\omega^b.\widehat{\nabla}_{X_b}\psi+\Omega.\psi)+(\omega^a)^{\eta}.\widehat{\nabla}_{X_a}(e^b.\widehat{\nabla}_{X_b}\psi)-\Omega^{\eta}.e^b.\widehat{\nabla}_{X_b}\psi
\nonumber\\
&=&i_{X a}\omega^b.\bigg(\widehat{\nabla}^2(X_a,X_b)+\widehat{\nabla}^2(X_b,X_a)\bigg)\psi+\bigg(e^b.\widehat{\nabla}_{X_b}\omega^a+2i_{X^a}\Omega\bigg).\widehat{\nabla}_{X_a}\psi\nonumber\\
&&+\bigg((e^a\wedge\omega^b).\widehat{R}(X_a,X_b)+e^a.\widehat{\nabla}_{X_a}\Omega\bigg).\psi
\end{eqnarray}
where we have used the following Clifford algebra identities
\begin{eqnarray}
e^a.\omega^b&=&e^a\wedge\omega^b+i_{X^a}\omega^b\nonumber\\
-(\omega^a)^{\eta}.e^b&=&-e^b\wedge\omega^a+i_{X^b}\omega^a\nonumber\\
e^a.\Omega&=&e^a\wedge\Omega+i_{X^a}\Omega\nonumber\\
-\Omega^{\eta}.e^a&=&-e^a\wedge\Omega+i_{X^a}\Omega\nonumber
\end{eqnarray}
and the definition of the Hessian
\begin{equation}
\widehat{\nabla}^2(X_a,X_b)=\widehat{\nabla}_{X_a}\widehat{\nabla}_{X_b}-\widehat{\nabla}_{\widehat{\nabla}_{X_a}X_b}
\end{equation}
which in normal coordinates reduces to $\widehat{\nabla}^2(X_a,X_b)=\widehat{\nabla}_{X_a}\widehat{\nabla}_{X_b}$.
The right hand side of (13) is written as
\begin{eqnarray}
M\widehat{\displaystyle{\not}D}\psi&=&2m^a.\widehat{\nabla}_{X_a}(e^b.\widehat{\nabla}_{X_b}\psi)+m.e^b.\widehat{\nabla}_{X_b}\psi\nonumber\\
&=&2m^a.e^b.\widehat{\nabla}^2(X_a,X_b)\psi+m.e^b.\widehat{\nabla}_{X_b}\psi.
\end{eqnarray}
By equating the second order terms in (13), we have
\begin{equation}
i_{X^a}\omega^b.(\widehat{\nabla}^2(X_a,X_b)+\widehat{\nabla}^2(X_b,X_a))\psi=2m^a.e^b.\widehat{\nabla}^2(X_a,X_b)\psi
\end{equation}
and by manipulating the indices we can write
\begin{equation}
i_{X^a}\omega^b+i_{X^b}\omega^a=m^a.e^b+m^b.e^a.
\end{equation}
Clifford multiplying (20) with $e_a$ from the right gives
\begin{eqnarray}
(i_{X^a}\omega^b+i_{X^b}\omega^a).e_a&=&m^a.e^b.e_a+m^b.e^a.e_a\nonumber\\
&=&(n+2)m^b-m^a.e_a.e^b
\end{eqnarray}
where we have used $e^a.e_a=n$ and $e^b.e_a+e_a.e^b=2g_a^b$. From (20), we have $i_{X^a}\omega_a=m^a.e_a$ and (21) can be written as
\begin{equation}
(n+2)m^b=(i_{X^a}\omega^b+i_{X^b}\omega^a).e_a+(i_{X^a}\omega_a).e^b
\end{equation}
and by multiplying with $e_b$ from the left
\begin{equation}
e_b.m^b=-i_{X_a}(\omega^a)^{\eta}.
\end{equation}
Similarly, if we multiply (20) with $e_a$ from the left, we can find
\begin{equation}
(n-2\pi)m^b=(\pi+1)(\omega^b)^{\eta}-i_{X^b}(e_a.(\omega^a)^{\eta})+(i_{X_a}\omega^a).e^b
\end{equation}
where $\pi$ acts on a $p$-form $\alpha$ as $\pi\alpha=p\alpha$. Adding (22) and (24) gives
\begin{equation}
(n+1-\pi)m^b=e^b\wedge (i_{X_a}\omega^a)^{\eta}
\end{equation}
and so (25) is a necessary condition for (20). In fact, we can arrange $m^b=0$ by adding a term to $L$ in (11) proportional to $\displaystyle{\not}D$ such as $\alpha.\displaystyle{\not}D$ which can be transferred to the right hand side of (13). If we add such a term to (11), then $\omega^a$ transforms into $\widehat{\omega}^a=\omega^a+\alpha.e^a$ and we can choose an $\alpha$ in terms of $\omega^a$ such as $m^b$ vanishes from (25). In that case, for the right hand side of (20) vanishes, the $p$-form $\omega^a$ can be written as the interior derivative of a $(p+1)$-form $\omega$ as follows
\begin{equation}
\omega^a=i_{X^a}\omega.
\end{equation} 
The details of this calculation can be found in \cite{Benn Kress, Acik Ertem Onder Vercin}. Equating the first order terms in (13) gives
\begin{eqnarray}
e^b.\widehat{\nabla}_{X_b}\omega^a+2i_{X^a}\Omega&=&m.e^a\nonumber\\
\implies e^b.i_{X^a}\widehat{\nabla}_{X_b}\omega+2i_{X^a}\Omega&=&m.e^a\nonumber\\
\implies \widehat{\nabla}_{X^a}\omega-i_{X^a}(e^b.\widehat{\nabla}_{X_b}\omega)+2i_{X^a}\Omega&=&m.e^a.
\end{eqnarray}
By taking the exterior product of (27) with $e_a$ gives
\begin{equation}
\widehat{d}\omega=\pi\mu
\end{equation}
where $\mu=\widehat{\displaystyle{\not}d}\omega-2\Omega-m^{\eta}$ and $\widehat{d}=e^a\wedge\widehat{\nabla}_{X_a}$ is the modified exterior derivative on differential forms. By contracting with $i_{X_a}$, we have for the $p$-form part
\begin{equation}
i_{X_a}\mu_{(p)}=\frac{1}{p+1}i_{X_a}\widehat{d}\omega_{(p)}
\end{equation}
where $\mu_{(p)}$ and $\omega_{(p)}$ are $p$-form parts. By contracting (27) with $i_{X_a}$, we can write
\begin{equation}
-\widehat{\delta}\omega=(n-\pi)m^{\eta}
\end{equation}
where $\widehat{\delta}=-i_{X^a}\widehat{\nabla}_{X_a}$ is the modified coderivative on differential forms and for the $p$-form part of the multiplication with $e_a$, we have
\begin{equation}
(e_a\wedge m^{\eta})_{(p)}=-\frac{1}{n-p+1}e_a\wedge\widehat{\delta}\omega_{(p)}.
\end{equation}
We can see that (27) is written for each $p$-form as
\begin{equation}
\widehat{\nabla}_{X^a}\omega=i_{X_a}\mu_{(p)}+(e_a\wedge m^{\eta})_{(p)}
\end{equation}
and we obtain by summing (29) and (31)
\begin{equation}
\widehat{\nabla}_{X_a}\omega_{(p)}=\frac{1}{p+1}i_{X_a}\widehat{d}\omega_{(p)}-\frac{1}{n-p+1}e_a\wedge\widehat{\delta}\omega_{(p)}.
\end{equation}
The details of the calculation can be found in \cite{Benn Kress, Acik Ertem Onder Vercin}. (33) corresponds to the modified conformal Killing-Yano (CKY) equation which is the generalization of the ordinary CKY equation written in terms of the unmodified connection and derivatives. So, $\omega^a$ in (11) have to be constructed from a modified CKY form $\omega$ satisfying (33) as in (26).

Equality of the zeroth order terms in (13) gives
\begin{equation}
(e^a\wedge i_{X^b}\omega).\widehat{R}(X_a,X_b)\psi+e^a.\widehat{\nabla}_{X_a}\Omega.\psi=0.
\end{equation}
By using the following identity
\begin{equation}
e^a\wedge i_{X^b}\omega=\frac{1}{4}\bigg(e^a.e^b.\omega-(-1)^pe^a.\omega.e^b-(-1)^pe^b.\omega.e^a+2g^{ab}\omega-\omega.e^a.e^b\bigg)
\end{equation}
and after doing some manipulations, we obtain
\begin{eqnarray}
&&\frac{1}{4}e^{ab}.\widehat{R}(X_a,X_b)(\omega.\psi)-\frac{1}{4}\omega.e^{ab}.\widehat{R}(X_a,X_b)\psi-\frac{1}{2}(e^a\wedge i_{X^b}\widehat{R}(X_a,X_b)\omega).\psi\nonumber\\
&&-\frac{1}{4}(e^{ab}\wedge\widehat{R}(X_a,X_b)\omega).\psi-\frac{1}{4}(i_{X^a}i_{X^b}\widehat{R}(X_a,X_b)\omega).\psi+e^a.\widehat{\nabla}_{X_a}\Omega.\psi=0.
\end{eqnarray}
We can transform wedge products and Clifford products into one another by Clifford algebra identities. For two differential forms $\alpha$ and $\beta$, we can write the wedge product in terms of the Clifford product as follows
\begin{equation}
\alpha\wedge\beta=\sum_{k=0}^{n}\frac{(-1)^{\lfloor k/2\rfloor}}{k!}(i_{X_{a_1}}...i_{X_{a_k}}\eta^{k}\alpha).i_{X^{a_1}}...i_{X^{a_k}}\beta.
\end{equation}
By using this identity, we can manipulate the curvature terms in (36) to write in terms of Clifford products as
\begin{equation}
e^{ab}\wedge\widehat{R}(X_a,X_b)\omega=e^{ab}.\widehat{R}(X_a,X_b)\omega-2e^a.i_{X^b}\widehat{R}(X_a,X_b)\omega+i_{X^a}i_{X^b}\widehat{R}(X_a,X_b)\omega
\end{equation}
and so we have
\begin{equation}
-\frac{1}{4}(e^{ab}\wedge\widehat{R}(X_a,X_b)\omega).\psi-\frac{1}{4}(i_{X^a}i_{X^b}\widehat{R}(X_a,X_b)\omega).\psi=-\frac{1}{4}e^{ab}.\widehat{R}(X_a,X_b)\omega.\psi+\frac{1}{2}(e^a\wedge i_{X^b}\widehat{R}(X_a,X_b)\omega).\psi.
\end{equation}
Then, (36) can be written as
\begin{equation}
\frac{1}{4}e^{ab}.\widehat{R}(X_a,X_b)(\omega.\psi)-\frac{1}{4}e^{ab}.\widehat{R}(X_a,X_b)\omega.\psi-\frac{1}{4}\omega.e^{ab}.\widehat{R}(X_a,X_b)\psi+e^a.\widehat{\nabla}_{X_a}\Omega.\psi=0.
\end{equation}
In fact, the second term in (40) has a special meaning in terms of modified CKY forms. It is defined as the curvature endomorphism operator
\begin{equation}
I(\widehat{R})\omega=e^{ab}.\widehat{R}(X_a,X_b)\omega
\end{equation}
and can be written as the integrability condition for the modified CKY equation. By taking the modified covariant derivative of (33) and doing some manipulations, one obtains the following integrability condition of the modified CKY equation \cite{Semmelman, Ertem, Acik Ertem}
\begin{equation}
\frac{p}{p+1}\widehat{\delta}\widehat{d}\omega+\frac{n-p}{n-p+1}\widehat{d}\widehat{\delta}\omega=-\frac{1}{2}e^{ab}.\widehat{R}(X_a,X_b)\omega.
\end{equation}

Now, we can consider the symmetry operator property of (11) for the modified Dirac operators of various types of modifications. Equating the second order and the first order terms in (13) will give the same conditions for all types of modifications. However, the equality of zeroth order terms in (13) will have different types of conditions in different cases. Let us consider the different types of modifications separately for the equality of zeroth order terms which corresponds to (40);

\textit{unmodified case:}\\
In that case, we can find the various curvature terms in (40) as follows. For the first term in (40), we have from (A52)
\begin{eqnarray}
\frac{1}{4}e^{ab}.R(X_a,X_b)(\omega.\psi)&=&\frac{1}{8}e^{ab}.R_{ab}.\omega.\psi\nonumber\\
&=&-\frac{1}{8}\mathcal{R}\omega.\psi.
\end{eqnarray}
For the second term, we can write from (42)
\begin{equation}
-\frac{1}{4}e^{ab}.R(X_a,X_b)\omega.\psi=\frac{p}{2(p+1)}\delta d\omega+\frac{n-p}{2(n-p+1)}d\delta\omega
\end{equation}
and for the third term, we obtain again from (A52)
\begin{eqnarray}
-\frac{1}{4}\omega.e^{ab}.R(X_a,X_b)\psi&=&-\frac{1}{8}\omega.e^{ab}.R_{ab}.\psi\nonumber\\
&=&\frac{1}{8}\mathcal{R}\omega.\psi.
\end{eqnarray}
The fourth term can be written as
\begin{equation}
e^a.\nabla_{X_a}\Omega.\psi=(d\Omega-\delta\Omega).\psi.
\end{equation}
By summing up the terms (43)-(46), we can write (40) as
\begin{equation}
\frac{p}{2(p+1)}\delta d\omega.\psi+\frac{n-p}{2(n-p+1)}d\delta\omega.\psi+(d\Omega-\delta\Omega).\psi=0
\end{equation}
and this will give a solution for $\Omega$ as
\begin{equation}
\Omega=\frac{p}{2(p+1)}d\omega-\frac{n-p}{2(n-p+1)}\delta\omega.
\end{equation}
Then, the symmetry operator (11) for the Dirac operator can be written in terms of CKY $p$-forms $\omega$ as
\begin{equation}
L=i_{X^a}\omega.\nabla_{X_a}+\frac{p}{2(p+1)}d\omega-\frac{n-p}{2(n-p+1)}\delta\omega.
\end{equation}

\textit{0-form modification:}\\
In that case, the first term in (40) will give from (A23) as
\begin{eqnarray}
\frac{1}{4}e^{ab}.\widehat{R}(X_a,X_b)(\omega.\psi)&=&\frac{1}{8}e^{ab}.R_{ab}.\omega.\psi+\frac{1}{2}k^2f^2e^{ab}.e_{ab}.\omega.\psi\nonumber\\
&&+\frac{1}{4}k(i_{X_a}df)e^{ab}.e_b.\omega.\psi-\frac{1}{4}k(i_{X_b}df)e^{ab}.e_a.\omega.\psi\nonumber\\
&=&\bigg(-\frac{1}{8}\mathcal{R}-\frac{n(n-2)}{2}k^2f^2+\frac{(n-1)}{2}kdf\bigg).\omega.\psi.
\end{eqnarray}
The second term in (40) will be
\begin{equation}
-\frac{1}{4}e^{ab}.\widehat{R}(X_a,X_b)\omega.\psi=\frac{p}{2(p+1)}\widehat{\delta}\widehat{d}\omega+\frac{n-p}{2(n-p+1)}\widehat{d}\widehat{\delta}\omega
\end{equation}
and the third term is
\begin{eqnarray}
-\frac{1}{4}\omega.e^{ab}.\widehat{R}(X_a,X_b)\psi&=&-\frac{1}{8}\omega.e^{ab}.R_{ab}.\psi-\frac{1}{2}k^2f^2e^{ab}.e_{ab}.\psi\nonumber\\
&&-\frac{1}{4}k(i_{X_a}df)\omega.e^{ab}.e_b.\psi+\frac{1}{4}k(i_{X_b}df)\omega.e^{ab}.e_a.\psi\nonumber\\
&=&\bigg(\frac{1}{8}\mathcal{R}+\frac{n(n-2)}{2}k^2f^2\bigg)\omega.\psi-\frac{(n-1)}{2}k\omega.df.\psi.
\end{eqnarray}
Hence, (40) will be written as
\begin{equation}
\frac{p}{2(p+1)}\widehat{\delta}\widehat{d}\omega.\psi+\frac{n-p}{2(n-p+1)}\widehat{d}\widehat{\delta}\omega.\psi+\frac{(n-1)}{2}k[df, \omega]_{Cl}.\psi+(\widehat{d}\Omega-\widehat{\delta}\Omega).\psi=0.
\end{equation}
Then, we have a solution for $\Omega$ as
\begin{equation}
\Omega=\frac{p}{2(p+1)}\widehat{d}\omega-\frac{n-p}{2(n-p+1)}\widehat{\delta}\omega
\end{equation}
with an extra condition on $\omega$ as
\begin{equation}
[df, \omega]_{Cl}=0.
\end{equation}
This means that we can construct a symmetry operator for 0-form modified Dirac operator  from 0-form modified CKY $p$-forms as in the following form
\begin{equation}
L=i_{X^a}\omega.\widehat{\nabla}_{X_a}+\frac{p}{2(p+1)}\widehat{d}\omega-\frac{n-p}{2(n-p+1)}\widehat{\delta}\omega.
\end{equation}
However, $\omega$ must also satisfy the extra condition (55). Namely, $\omega$ must Clifford commute with the exterior derivative of the function $f$ appearing in the modification of the Dirac operator. Only in that case the operator $L$ in (56) will be a symmetry operator for the 0-form modified Dirac operator.

\textit{1-form modification:}\\
For the 1-form modified case, the curvature terms can be calculated by using (A25) and found as
\begin{eqnarray}
\frac{1}{4}e^{ab}.\widehat{R}(X_a,X_b)(\omega.\psi)&=&-\frac{1}{8}\mathcal{R}\omega.\psi+\frac{k_2-(n-3)k_1}{2}dA.\omega.\psi-\frac{(n-1)}{2}(\delta A)\omega.\psi\nonumber\\
&&+\frac{k_1^2}{2}\bigg(n(n-3)-2(n-2)\bigg)(A\wedge A).\omega.\psi+\frac{k_1^2}{2}\bigg(n(n-3)+2\bigg)(A\underset{1}{\wedge}A)\omega.\psi
\end{eqnarray}
and
\begin{eqnarray}
-\frac{1}{4}\omega.e^{ab}.\widehat{R}(X_a,X_b)\psi&=&\frac{1}{8}\mathcal{R}\omega.\psi-\frac{k_2-(n-3)k_1}{2}\omega.dA.\psi+\frac{(n-1)}{2}k_1(\delta A)\omega.\psi\nonumber\\
&&-\frac{k_1^2}{2}\bigg(n(n-3)-2(n-2)\bigg)\omega.(A\wedge A).\psi-\frac{k_1^2}{2}\bigg(n(n-3)+2\bigg)(A\underset{1}{\wedge}A)\omega.\psi.
\end{eqnarray}
where the contracted wedge product $\underset{k}{\wedge}$ is defined in (A42). The other curvature term is in the same form as in (51). Then, (40) is written as
\begin{eqnarray}
&&\frac{p}{2(p+1)}\widehat{\delta}\widehat{d}\omega.\psi+\frac{n-p}{2(n-p+1)}\widehat{d}\widehat{\delta}\omega.\psi+\frac{k_2-(n-3)k_1}{2}[dA, \omega]_{Cl}.\psi\nonumber\\
&&+\frac{k_1^2}{2}\bigg(n(n-3)-2(n-2)\bigg)[A\wedge A, \omega]_{Cl}.\psi+(\widehat{d}\Omega-\widehat{\delta}\Omega).\psi=0.
\end{eqnarray}
Then, we have a solution for $\Omega$ as
\begin{equation}
\Omega=\frac{p}{2(p+1)}\widehat{d}\omega-\frac{n-p}{2(n-p+1)}\widehat{\delta}\omega
\end{equation}
with an extra condition on $\omega$
\begin{equation}
\frac{k_2-(n-3)k_1}{2}[dA, \omega]_{Cl}+\frac{k_1^2}{2}\bigg(n(n-3)-2(n-2)\bigg)[A\wedge A, \omega]_{Cl}=0
\end{equation}
or two different extra conditions on $\omega$ for $k_1\neq 0$ and $k_2\neq 0$
\begin{eqnarray}
{[dA, \omega]_{Cl}}&=&0\\
{[A\wedge A, \omega]_{Cl}}&=&0.
\end{eqnarray}
This means that we can construct a symmetry operator for 1-form modified Dirac operator from 1-form modified CKY $p$-forms as in the following form
\begin{equation}
L=i_{X^a}\omega.\widehat{\nabla}_{X_a}+\frac{p}{2(p+1)}\widehat{d}\omega-\frac{n-p}{2(n-p+1)}\widehat{\delta}\omega.
\end{equation}
and $\omega$ must also satisfy the extra condition (61) or both the conditions (62) and (63). For the special choice of $k_1=0$ and $k_2=1$, only the condition (62) on $\omega$ remains and this corresponds to the symmetry operators of Dirac operators in the presence of minimal coupling which is demonstrated in \cite{Acik Ertem Onder Vercin}.

\textit{2-form modification:}\\
For the 2-form modified case, by using (A28) the curvature terms can be calculated as
\begin{eqnarray}
\frac{1}{4}e^{ab}.\widehat{R}(X_a,X_b)(\omega.\psi)&=&-\frac{1}{8}\mathcal{R}\omega.\psi+\frac{(n-5)k_1+2k_2}{2}dF.\omega.\psi-\frac{(n-3)k_1+k_2}{2}\delta F.\omega.\psi\nonumber\\
&&-\frac{1}{2}\bigg((n-4)(n-9)k_1^2+4(n-6)k_1k_2+4k_2^2\bigg)(F\wedge F).\omega.\psi\nonumber\\
&&+\frac{1}{2}\bigg((n-4)(n-5)k_1^2-2k_1k_2-3k_2^2\bigg)(F\underset{1}{\wedge}F).\omega.\psi\nonumber\\
&&+\frac{1}{2}\bigg(\frac{(n-1)(n-4)}{2}k_1^2+2k_1k_2-k_2^2\bigg)(F\underset{2}{\wedge}F)\omega.\psi
\end{eqnarray}
and
\begin{eqnarray}
-\frac{1}{4}\omega.e^{ab}.\widehat{R}(X_a,X_b)\psi&=&\frac{1}{8}\mathcal{R}\omega.\psi-\frac{(n-5)k_1+2k_2}{2}\omega.dF.\psi+\frac{(n-3)k_1+k_2}{2}\omega.\delta F.\psi\nonumber\\
&&+\frac{1}{2}\bigg((n-4)(n-9)k_1^2+4(n-6)k_1k_2+4k_2^2\bigg)\omega.(F\wedge F).\psi\nonumber\\
&&-\frac{1}{2}\bigg((n-4)(n-5)k_1^2-2k_1k_2-3k_2^2\bigg)\omega.(F\underset{1}{\wedge}F).\psi\nonumber\\
&&-\frac{1}{2}\bigg(\frac{(n-1)(n-4)}{2}k_1^2+2k_1k_2-k_2^2\bigg)(F\underset{2}{\wedge}F)\omega.\psi.
\end{eqnarray}
The other curvature term is in the same form as in (51). Then, (40) is written as
\begin{eqnarray}
&&\frac{p}{2(p+1)}\widehat{\delta}\widehat{d}\omega.\psi+\frac{n-p}{2(n-p+1)}\widehat{d}\widehat{\delta}\omega.\psi+\frac{(n-5)k_1+2k_2}{2}[dF, \omega]_{Cl}.\psi\nonumber\\
&&-\frac{(n-3)k_1+k_2}{2}[\delta F, \omega]_{Cl}.\psi-\frac{1}{2}\bigg((n-4)(n-9)k_1^2+4(n-6)k_1k_2+4k_2^2\bigg)[F\wedge F, \omega]_{Cl}.\psi\nonumber\\
&&+\frac{1}{2}\bigg((n-4)(n-5)k_1^2-2k_1k_2-3k_2^2\bigg)[F\underset{1}{\wedge}F, \omega]_{Cl}.\psi+(\widehat{d}\Omega-\widehat{\delta}\Omega).\psi=0.
\end{eqnarray}
Then, we have a solution for $\Omega$ as
\begin{equation}
\Omega=\frac{p}{2(p+1)}\widehat{d}\omega-\frac{n-p}{2(n-p+1)}\widehat{\delta}\omega
\end{equation}
with an extra condition on $\omega$
\begin{eqnarray}
&&\frac{(n-5)k_1+2k_2}{2}[dF, \omega]_{Cl}-\frac{(n-3)k_1+k_2}{2}[\delta F, \omega]_{Cl}\nonumber\\
&&-\frac{1}{2}\bigg((n-4)(n-9)k_1^2+4(n-6)k_1k_2+4k_2^2\bigg)[F\wedge F, \omega]_{Cl}\nonumber\\
&&+\frac{1}{2}\bigg((n-4)(n-5)k_1^2-2k_1k_2-3k_2^2\bigg)[F\underset{1}{\wedge}F, \omega]_{Cl}=0
\end{eqnarray}
or four different extra conditions on $\omega$ for $k_1\neq 0$ and $k_2\neq 0$
\begin{eqnarray}
{[dF, \omega]_{Cl}}&=&0\\
{[\delta F, \omega]_{Cl}}&=&0\\
{[F\wedge F, \omega]_{Cl}}&=&0\\
{[F\underset{1}{\wedge} F, \omega]_{Cl}}&=&0.
\end{eqnarray}
This means that we can construct a symmetry operator for 2-form modified Dirac operator from 2-form modified CKY $p$-forms as in the following form
\begin{equation}
L=i_{X^a}\omega.\widehat{\nabla}_{X_a}+\frac{p}{2(p+1)}\widehat{d}\omega-\frac{n-p}{2(n-p+1)}\widehat{\delta}\omega.
\end{equation}
and $\omega$ must also satisfy the extra condition (69) or all of the conditions (70-73).

\textit{3-form modification:}\\
For the 3-form modified case, the curvature terms can be calculated from (A36) as
\begin{eqnarray}
\frac{1}{4}e^{ab}.\widehat{R}(X_a,X_b)(\omega.\psi)&=&-\frac{1}{8}\mathcal{R}\omega.\psi+\frac{3k_2-(n-7)k_1}{2}dH.\omega.\psi+\frac{(n-5)k_1+2k_2}{2}\delta H.\omega.\psi\nonumber\\
&&+\frac{1}{2}\bigg((n-6)(n-13)k_1^2+18k_1k_2-9k_2^2\bigg)(H\wedge H).\omega.\psi\nonumber\\
&&+\frac{1}{2}\bigg((n-6)(n-9)k_1^2-4(n-8)k_1k_2+4k_2^2\bigg)(H\underset{1}{\wedge}H).\omega.\psi\nonumber\\
&&+\frac{1}{2}\bigg(\frac{(n-6)(n-5)}{2}k_1^2+k_1k_2+\frac{k_2^2}{2}\bigg)(H\underset{2}{\wedge}H).\omega.\psi\nonumber\\
&&+\frac{1}{2}\bigg(-\frac{(n-1)(n-6)}{6}k_1^2+2k_1k_2+k_2^2\bigg)(H\underset{3}{\wedge}H)\omega.\psi
\end{eqnarray}
and
\begin{eqnarray}
-\frac{1}{4}\omega.e^{ab}.\widehat{R}(X_a,X_b)\psi&=&\frac{1}{8}\mathcal{R}\omega.\psi-\frac{3k_2-(n-7)k_1}{2}\omega.dH.\psi-\frac{(n-5)k_1+2k_2}{2}\omega.\delta H.\psi\nonumber\\
&&-\frac{1}{2}\bigg((n-6)(n-13)k_1^2+18k_1k_2-9k_2^2\bigg)\omega.(H\wedge H).\psi\nonumber\\
&&-\frac{1}{2}\bigg((n-6)(n-9)k_1^2-4(n-8)k_1k_2+4k_2^2\bigg)\omega.(H\underset{1}{\wedge}H).\psi\nonumber\\
&&-\frac{1}{2}\bigg(-\frac{(n-5)(n-6)}{2}k_1^2+k_1k_2+\frac{k_2^2}{2}\bigg)\omega.(H\underset{2}{\wedge}H).\psi\nonumber\\
&&-\frac{1}{2}\bigg(-\frac{(n-1)(n-6)}{6}k_1^2+2k_1k_2+k_2^2\bigg)(H\underset{3}{\wedge}H)\omega.\psi.
\end{eqnarray}
The other curvature term is in the same form as in (51). Then, (40) is written as
\begin{eqnarray}
&&\frac{p}{2(p+1)}\widehat{\delta}\widehat{d}\omega.\psi+\frac{n-p}{2(n-p+1)}\widehat{d}\widehat{\delta}\omega.\psi+\frac{3k_2-(n-7)k_1}{2}[dH, \omega]_{Cl}.\psi\nonumber\\
&&+\frac{(n-5)k_1+2k_2}{2}[\delta H, \omega]_{Cl}.\psi+\frac{1}{2}\bigg((n-6)(n-13)k_1^2+18k_1k_2-9k_2^2\bigg)[H\wedge H, \omega]_{Cl}.\psi\nonumber\\
&&+\frac{1}{2}\bigg((n-6)(n-9)k_1^2-4(n-8)k_1k_2+4k_2^2\bigg)[H\underset{1}{\wedge}H, \omega]_{Cl}.\psi\nonumber\\
&&+\frac{1}{2}\bigg(-\frac{(n-5)(n-6)}{2}k_1^2+k_1k_2+\frac{k_2^2}{2}\bigg)[H\underset{2}{\wedge}H, \omega]_{Cl}.\psi+(\widehat{d}\Omega-\widehat{\delta}\Omega).\psi=0.
\end{eqnarray}
Then, we have a solution for $\Omega$ as
\begin{equation}
\Omega=\frac{p}{2(p+1)}\widehat{d}\omega-\frac{n-p}{2(n-p+1)}\widehat{\delta}\omega
\end{equation}
with an extra condition on $\omega$
\begin{eqnarray}
&&\frac{3k_2-(n-7)k_1}{2}[dH, \omega]_{Cl}+\frac{(n-5)k_1+2k_2}{2}[\delta H, \omega]_{Cl}\nonumber\\
&&+\frac{1}{2}\bigg((n-6)(n-13)k_1^2+18k_1k_2-9k_2^2\bigg)[H\wedge H, \omega]_{Cl}\nonumber\\
&&+\frac{1}{2}\bigg((n-6)(n-9)k_1^2-4(n-8)k_1k_2+4k_2^2\bigg)[H\underset{1}{\wedge}H, \omega]_{Cl}\nonumber\\
&&+\frac{1}{2}\bigg(-\frac{(n-5)(n-6)}{2}k_1^2+k_1k_2+\frac{k_2^2}{2}\bigg)[H\underset{2}{\wedge}H, \omega]_{Cl}=0
\end{eqnarray}
or five different extra conditions on $\omega$ for $k_1\neq 0$ and $k_2\neq 0$
\begin{eqnarray}
{[dH, \omega]_{Cl}}&=&0\\
{[\delta H, \omega]_{Cl}}&=&0\\
{[H\wedge H, \omega]_{Cl}}&=&0\\
{[H\underset{1}{\wedge} H, \omega]_{Cl}}&=&0\\
{[H\underset{2}{\wedge} H, \omega]_{Cl}}&=&0.
\end{eqnarray}
This means that we can construct a symmetry operator for 3-form modified Dirac operator from 3-form modified CKY $p$-forms as in the following form
\begin{equation}
L=i_{X^a}\omega.\widehat{\nabla}_{X_a}+\frac{p}{2(p+1)}\widehat{d}\omega-\frac{n-p}{2(n-p+1)}\widehat{\delta}\omega.
\end{equation}
and $\omega$ must also satisfy the extra condition (79) or all of the conditions (80-84). For $k_1=0$ and $k_2=\frac{1}{4}$, (85) corresponds to the symmetry operator for 3-form modified Dirac operators on 10-dimensional supergravity backgrounds with 3-form flux $H$.

\textit{4-form modification:}\\
For the 4-form modified case, the curvature terms can be calculated from (A44) as
\begin{eqnarray}
\frac{1}{4}e^{ab}.\widehat{R}(X_a,X_b)(\omega.\psi)&=&-\frac{1}{8}\mathcal{R}\omega.\psi+\left(\frac{n-9}{2}k_1+2k_2\right)dG.\omega.\psi-\frac{(n-7)k_1+3k_2}{2}\delta G.\omega.\psi\nonumber\\
&&+\bigg(-\frac{(n-8)(n-15)}{2}k_1^2-4(n-12)k_1k_2-8k_2^2\bigg)(G\wedge G).\omega.\psi\nonumber\\
&&+\bigg(\frac{(n-8)(n-11)}{2}k_1^2-9k_1k_2-\frac{15}{2}k_2^2\bigg)(G\underset{1}{\wedge}G).\omega.\psi\nonumber\\
&&+\bigg(\frac{(n-8)(n-7)}{4}k_1^2+(n-9)k_1k_2+\frac{k_2^2}{2}\bigg)(G\underset{2}{\wedge}G).\omega.\psi\nonumber\\
&&+\bigg(-\frac{(n-8)(n-5)}{3}k_1^2-\frac{k_1k_2}{2}+\frac{13}{12}k_2^2\bigg)(G\underset{3}{\wedge}G).\omega.\psi\nonumber\\
&&+\bigg(-\frac{(n-8)(n-1)}{12}k_1^2-\frac{k_1k_2}{2}+\frac{k_2^2}{4}\bigg)(G\underset{4}{\wedge}G)\omega.\psi
\end{eqnarray}
and
\begin{eqnarray}
-\frac{1}{4}\omega.e^{ab}.\widehat{R}(X_a,X_b)\psi&=&\frac{1}{8}\mathcal{R}\omega.\psi-\left(\frac{n-9}{2}k_1+2k_2\right)\omega.dG.\psi+\frac{(n-7)k_1+3k_2}{2}\omega.\delta G.\psi\nonumber\\
&&-\bigg(-\frac{(n-8)(n-15)}{2}k_1^2-4(n-12)k_1k_2-8k_2^2\bigg)\omega.(G\wedge G).\psi\nonumber\\
&&-\bigg(\frac{(n-8)(n-11)}{2}k_1^2-9k_1k_2-\frac{15}{2}k_2^2\bigg)\omega.(G\underset{1}{\wedge}G).\psi\nonumber\\
&&-\bigg(\frac{(n-8)(n-7)}{4}k_1^2+(n-9)k_1k_2+\frac{k_2^2}{2}\bigg)\omega.(G\underset{2}{\wedge}G).\psi\nonumber\\
&&-\bigg(-\frac{(n-8)(n-5)}{3}k_1^2-\frac{k_1k_2}{2}+\frac{13}{12}k_2^2\bigg)\omega.(G\underset{3}{\wedge}G).\psi\nonumber\\
&&-\bigg(-\frac{(n-8)(n-1)}{12}k_1^2-\frac{k_1k_2}{2}+\frac{k_2^2}{4}\bigg)(G\underset{4}{\wedge}G)\omega.\psi.
\end{eqnarray}
The other curvature term is in the same form as in (51). Then, (40) is written as
\begin{eqnarray}
&&\frac{p}{2(p+1)}\widehat{\delta}\widehat{d}\omega.\psi+\frac{n-p}{2(n-p+1)}\widehat{d}\widehat{\delta}\omega.\psi+\left(\frac{n-9}{2}k_1+2k_2\right)[dG, \omega]_{Cl}.\psi\nonumber\\
&&-\frac{(n-7)k_1+3k_2}{2}[\delta G, \omega]_{Cl}.\psi+\bigg(-\frac{(n-8)(n-15)}{2}k_1^2-4(n-12)k_1k_2-8k_2^2\bigg)[G\wedge G, \omega]_{Cl}.\psi\nonumber\\
&&+\bigg(\frac{(n-8)(n-11)}{2}k_1^2-9k_1k_2-\frac{15}{2}k_2^2\bigg)[G\underset{1}{\wedge}G, \omega]_{Cl}.\psi\nonumber\\
&&+\bigg(\frac{(n-8)(n-7)}{4}k_1^2+(n-9)k_1k_2+\frac{k_2^2}{2}\bigg)[G\underset{2}{\wedge}G, \omega]_{Cl}.\psi\nonumber\\
&&+\bigg(-\frac{(n-8)(n-5)}{3}k_1^2-\frac{k_1k_2}{2}+\frac{13}{12}k_2^2\bigg)[G\underset{3}{\wedge}G, \omega]_{Cl}.\psi+(\widehat{d}\Omega-\widehat{\delta}\Omega).\psi=0.
\end{eqnarray}
Then, we have a solution for $\Omega$ as
\begin{equation}
\Omega=\frac{p}{2(p+1)}\widehat{d}\omega-\frac{n-p}{2(n-p+1)}\widehat{\delta}\omega
\end{equation}
with an extra condition on $\omega$
\begin{eqnarray}
&&\left(\frac{n-9}{2}k_1+2k_2\right)[dG, \omega]_{Cl}-\frac{(n-7)k_1+3k_2}{2}[\delta G, \omega]_{Cl}\nonumber\\
&&+\bigg(-\frac{(n-8)(n-15)}{2}k_1^2-4(n-12)k_1k_2-8k_2^2\bigg)[G\wedge G, \omega]_{Cl}\nonumber\\
&&+\bigg(\frac{(n-8)(n-11)}{2}k_1^2-9k_1k_2-\frac{15}{2}k_2^2\bigg)[G\underset{1}{\wedge}G, \omega]_{Cl}\nonumber\\
&&+\bigg(\frac{(n-8)(n-7)}{4}k_1^2+(n-9)k_1k_2+\frac{k_2^2}{2}\bigg)[G\underset{2}{\wedge}G, \omega]_{Cl}\nonumber\\
&&+\bigg(-\frac{(n-8)(n-5)}{3}k_1^2-\frac{k_1k_2}{2}+\frac{13}{12}k_2^2\bigg)[G\underset{3}{\wedge}G, \omega]_{Cl}=0
\end{eqnarray}
or six different extra conditions on $\omega$ for $k_1\neq 0$ and $k_2\neq 0$
\begin{eqnarray}
{[dG, \omega]_{Cl}}&=&0\\
{[\delta G, \omega]_{Cl}}&=&0\\
{[G\wedge G, \omega]_{Cl}}&=&0\\
{[G\underset{1}{\wedge} G, \omega]_{Cl}}&=&0\\
{[G\underset{2}{\wedge} G, \omega]_{Cl}}&=&0\\
{[G\underset{3}{\wedge} G, \omega]_{Cl}}&=&0.
\end{eqnarray}
This means that we can construct a symmetry operator for 4-form modified Dirac operator from 4-form modified CKY $p$-forms as in the following form
\begin{equation}
L=i_{X^a}\omega.\widehat{\nabla}_{X_a}+\frac{p}{2(p+1)}\widehat{d}\omega-\frac{n-p}{2(n-p+1)}\widehat{\delta}\omega.
\end{equation}
and $\omega$ must also satisfy the extra condition (90) or all of the conditions (91-96). For $k_1=-\frac{1}{12}$ and $k_2=\frac{1}{12}$, (97) corresponds to the symmetry operator for 4-form modified Dirac operators on 11-dimensional supergravity backgrounds with 4-form flux $G$.

\section{Symmetry Operators for Massive Modified Dirac Operators}

We consider the modified Dirac equation with mass $m$ for a spinor $\psi$
\begin{equation}
\widehat{\displaystyle{\not}D}\psi=m\psi.
\end{equation}
For the operator $L$ defined in (11) to be a symmetry operator for the massive modified Dirac equation, it has to satisfy the following condition
\begin{equation}
[\widehat{\displaystyle{\not}D}, L]_{gCl}=0.
\end{equation}
In that case, if $\psi$ is a solution of the massive modified Dirac equation, then $L\psi$ is also a solution for it.

The left hand side of (99) can be calculated as in (16)
\begin{eqnarray}
[\widehat{\displaystyle{\not}D}, L]_{gCl}\psi&=&\widehat{\displaystyle{\not}D}L\psi-L^{\eta}\widehat{\displaystyle{\not}D}\psi\nonumber\\
&=&i_{X a}\omega^b.\bigg(\widehat{\nabla}^2(X_a,X_b)+\widehat{\nabla}^2(X_b,X_a)\bigg)\psi+\bigg(e^b.\widehat{\nabla}_{X_b}\omega^a+2i_{X^a}\Omega\bigg).\widehat{\nabla}_{X_a}\psi\nonumber\\
&&+\bigg((e^a\wedge\omega^b).\widehat{R}(X_a,X_b)+e^a.\widehat{\nabla}_{X_a}\Omega\bigg).\psi.
\end{eqnarray}
While the second order terms in (99) give
\begin{equation}
\omega^a=i_{X^a}\omega
\end{equation}
the first order terms give the following equation
\begin{equation}
\widehat{\nabla}_{X_a}\omega_{(p)}=\frac{1}{p+1}i_{X_a}\widehat{d}\omega_{(p)}
\end{equation}
where $\omega_{(p)}$ is the $p$-form. (102) is the modified KY equation which is the generalization of the ordinary KY equation written in terms of the unmodified connection and exterior derivative. So $\omega^a$ in (11) must be constructed from a modified KY form $\omega$ for $L$ be a symmetry operator for the massive modified Dirac equation.
The analysis of zeroth order terms is completely equivalent to the massless case and the same conditions in terms of fluxes appear as in the massless case to construct a symmetry operator for massive modified Dirac equation in terms of modified KY forms. 

\section{Conclusion}

By using the modified Dirac operators which are written in terms of various degrees of fluxes, we construct modified curvature operators and obtain Schr\"odinger-Lichnerowicz-like formulas for squares of modified Dirac operators. From this machinery, we obtain the symmetry operators of massless and massive modified Dirac operators in terms of modified KY and CKY forms. Moreover, we obtain extra constraints to construct symmetry operators in terms of fluxes that modifiy the Dirac operators. Our results are relevant for all types of fluxes in all dimensions. However, for special types of fluxes in special dimensions, we find the symmetry operators of modified Dirac operators in 5-, 6-, 10- and 11-dimensional supergravity theories.

From another point of view, the construction of symmetry operators for modified Dirac operators give a meaning to the modification of hidden symmetries in terms of fluxes. Modified KY and CKY forms are defined in terms of fluxes and the constraints give some subclasses of modified hidden symmetries. So, the algebraic structure of modified hidden symmetries and their subclasses can be investigated to obtain more general structures on supergravity flux backgrounds which can give different types of classification properties for supergravity backgrounds in various dimensions.

\begin{acknowledgments}

This study was supported by Scientific and Technological Research Council of T\"urkiye (T\"UB\.ITAK) under the Grant Number 123F261. The author thanks to T\"UB\.ITAK for their supports.

\end{acknowledgments}

\begin{appendix}

\section{Explicit calculations of modified operators for special cases}

In this Appendix, we consider the explicit calculations of modified Dirac operators, spin curvature operators and squares of the Dirac operators for the special cases of $p=$0, 1, 2, 3 and 4 which are important in supergravity applications. 

\subsection{Modified spin connections and Dirac operators}

For a function or a 0-form $f$, we can write the 0-form modified spin connection as
\begin{equation}
\widehat{\nabla}_X\psi = \nabla_X\psi+kf\widetilde{X}.\psi
\end{equation}
where $k$ is a constant. So, the 0-form modified Dirac operator is
\begin{eqnarray}
\widehat{\displaystyle{\not}D}\psi&=&\displaystyle{\not}D\psi+kfe^a.e_a.\psi\nonumber\\
&=&\displaystyle{\not}D\psi+nkf\psi
\end{eqnarray}
where we have used the Clifford algebra identity $e^a.e_a=n$.

For a 1-form $A$, we have the 1-form modified spin connection
\begin{equation}
\widehat{\nabla}_X\psi=\nabla_X\psi+k_1A.\widetilde{X}.\psi+k_2(i_XA)\psi
\end{equation}
and the 1-form modified Dirac operator
\begin{eqnarray}
\widehat{\displaystyle{\not}D}\psi&=&\displaystyle{\not}D\psi+(k_2-(n-2)k_1)A.\psi
\end{eqnarray}
For the special choices of $k_1=0$ and $k_2=1$, the modified spin connection and Dirac operator correspond to the following equalities
\begin{eqnarray}
\widehat{\nabla}_X\psi&=&\nabla_X\psi+(i_XA)\psi\\
\widehat{\displaystyle{\not}D}\psi&=&\displaystyle{\not}D\psi+A.\psi
\end{eqnarray}
which correspond to the gauged spin connection and gauged Dirac operator defined on Spin$^c$ manifolds \cite{Ertem2}.

For a 2-form $F$, the 2-form modified spin connection is
\begin{equation}
\widehat{\nabla}_X\psi=\nabla_X\psi+k_1F.\widetilde{X}.\psi+k_2(i_XF).\psi
\end{equation}
and the 2-form modified Dirac operator is
\begin{eqnarray}
\widehat{\displaystyle{\not}D}\psi&=&\displaystyle{\not}D\psi+((n-4)k_1+2k_2)F.\psi.
\end{eqnarray}
For the special choice of $k_1=\frac{1}{2\sqrt{3}}$ and $k_2=-\frac{1}{2\sqrt{3}}$, (10) and (11) reduce to the modified spin connection and Dirac operator defined in five dimensional supergravity with respect to a flux 2-form $F$
\begin{eqnarray}
\widehat{\nabla}_X\psi&=&\nabla_X\psi+\frac{1}{2\sqrt{3}}\left(F.\widetilde{X}-i_XF\right).\psi\\
\widehat{\displaystyle{\not}D}\psi&=&\displaystyle{\not}D\psi-\frac{1}{2\sqrt{3}}F.\psi.
\end{eqnarray}

For a 3-form $H$, we can write the 3-form modified spin connection as
\begin{equation}
\widehat{\nabla}_X\psi=\nabla_X\psi+k_1H.\widetilde{X}.\psi+k_2(i_XH).\psi
\end{equation}
and the corresponding 3-form modified Dirac operator as
\begin{eqnarray}
\widehat{\displaystyle{\not}D}\psi&=&\displaystyle{\not}D\psi+(3k_2-(n-6)k_1)H.\psi.
\end{eqnarray}
For the special choice of $k_1=0$ and $k_2=\frac{1}{4}$, (A11) and (A12) reduce to the modified spin connection and Dirac operator defined in the six or ten dimensional supergravity theories in terms of the flux 3-form $H$
\begin{eqnarray}
\widehat{\nabla}_X\psi&=&\nabla_X\psi+\frac{1}{4}i_XH.\psi\\
\widehat{\displaystyle{\not}D}\psi&=&\displaystyle{\not}D\psi+\frac{3}{4}H.\psi.
\end{eqnarray}

For a 4-form $G$, the 4-form modified spin connection is written as
\begin{equation}
\widehat{\nabla}_X\psi=\nabla_X\psi+k_1G.\widetilde{X}.\psi+k_2(i_XG)\psi
\end{equation}
and the 4-form modified Dirac operator is
\begin{eqnarray}
\widehat{\displaystyle{\not}D}\psi&=&\displaystyle{\not}D\psi+((n-8)k_1+4k_2)G.\psi.
\end{eqnarray}
If we use the relation between Clifford product and wedge product
\begin{equation}
G.\widetilde{X}=\widetilde{X}\wedge G-i_XG
\end{equation}
we can write the 4-form modified spin connection as
\begin{eqnarray}
\widehat{\nabla}_X\psi&=&\nabla_X\psi+k_1(\widetilde{X}\wedge G-i_XG).\psi+k_2(i_XG).\psi\nonumber\\
&=&\nabla_X\psi+k_1(\widetilde{X}\wedge G).\psi+(k_2-k_1)i_XG.\psi
\end{eqnarray}
and if we choose $k_1=-\frac{1}{12}$ and $k_2=\frac{1}{12}$, we arrive at the modified spin connection
\begin{equation}
\widehat{\nabla}_X\psi=\nabla_X\psi-\frac{1}{12}(\widetilde{X}\wedge G).\psi+\frac{1}{6}i_XG.\psi
\end{equation}
which arises in eleven dimensional supergravity. In that case the modified Dirac operator is written for $n=11$ as
\begin{equation}
\widehat{\displaystyle{\not}D}\psi=\displaystyle{\not}D\psi+\frac{1}{12}G.\psi.
\end{equation}

\subsection{Spin curvature operators for modified spin connections}

We consider various cases of different degree form modifications below.

\textit{0-form modification}:

From (7), we can write the 0-form modified spin curvature operator as
\begin{eqnarray}
\widehat{R}(X,Y)\psi&=&R(X,Y)\psi+k(X(f)\widetilde{Y}-Y(f)\widetilde{X}).\psi+kf(\nabla_X\widetilde{Y}-\nabla_Y\widetilde{X}).\psi\nonumber\\
&&+k^2f^2[\widetilde{X}, \widetilde{Y}]_{Cl}.\psi-kf\widetilde{[X, Y]}.\psi
\end{eqnarray}
where $X(f)$ denotes the derivative of $f$ along $X$. From the Clifford algebra identities
\begin{eqnarray}
\widetilde{X}.\widetilde{Y}&=&\widetilde{X}\wedge \widetilde{Y}+i_X\widetilde{Y}\nonumber\\
\widetilde{Y}.\widetilde{X}&=&-\widetilde{X}\wedge \widetilde{Y}+i_X\widetilde{Y}
\end{eqnarray}
we can write $[\widetilde{X}, \widetilde{Y}]_{Cl}=2\widetilde{X}\wedge \widetilde{Y}$. So, in normal coordinates we obtain the action of the modified spin curvature operator on a spinor as follows
\begin{eqnarray}
\widehat{R}(X_a,X_b)\psi&=&R(X_a,X_b)\psi+k(X_a(f)e_b-X_b(f)e_a).\psi+2k^2f^2e_{ab}.\psi\nonumber\\
&=&\left(\frac{1}{2}R_{ab}+2k^2f^2e_{ab}+k((i_{X_a}df)e_b-(i_{X_b}df)e_a)\right).\psi
\end{eqnarray}
where we have used the abbreviation $e_{ab}=e_a\wedge e_b$, $X_a(f)=i_{X_a}df$ and (5).

\textit{1-form modification}:

The 1-form modified spin curvature operator can be written as
\begin{eqnarray}
\widehat{R}(X,Y)\psi&=&R(X,Y)\psi+k_1(\nabla_XA.\widetilde{Y}-\nabla_YA.\widetilde{X}).\psi\nonumber\\
&&+k_1A.(\nabla_X\widetilde{Y}-\nabla_Y\widetilde{X}).\psi+k_2(\nabla_Xi_YA-\nabla_Yi_XA).\psi\nonumber\\
&&+k_1^2A.(\widetilde{X}.A.\widetilde{Y}-\widetilde{Y}.A.\widetilde{X}).\psi-k_1A.\widetilde{[X,Y]}.\psi-k_2(i_{[X,Y]}A)\psi
\end{eqnarray}
and by using normal coordinates, we can write
\begin{eqnarray}
\widehat{R}(X_a,X_b)\psi&=&R(X_a,X_b)\psi+k_1(\nabla_{X_a}A.e_b-\nabla_{X_b}A.e_a).\psi\nonumber\\
&&-k_2(i_{X_a}i_{X_b}dA)\psi+k_1^2A.(e_a.A.e_b-e_b.A.e_a).\psi
\end{eqnarray}
where we have used $i_{X_a}i_{X_b}dA=\nabla_{X_b}i_{X_a}A-\nabla_{X_a}i_{X_b}A$ in normal coordinates. For $k_1=0$ and $k_2=1$, we obtain the action of the spin curvature operator for gauged spin connection
\begin{equation}
\widehat{R}(X_a,X_b)\psi=R(X_a,X_b)\psi-(i_{X_a}i_{X_b}dA)\psi.
\end{equation}

\textit{2-form modification}:

For the 2-form modified spin curvature operator, we have
\begin{eqnarray}
\widehat{R}(X,Y)\psi&=&R(X,Y)\psi+k_1(\nabla_XF.\widetilde{Y}-\nabla_YF.\widetilde{X}).\psi\nonumber\\
&&+k_1F.(\nabla_X\widetilde{Y}-\nabla_Y\widetilde{X}).\psi+k_2(\nabla_Xi_YF-\nabla_Yi_XF).\psi\nonumber\\
&&+k_1^2F.(\widetilde{X}.F.\widetilde{Y}-\widetilde{Y}.F.\widetilde{X}).\psi+k_1k_2[F.\widetilde{X}, i_YF]_{Cl}.\psi+k_1k_2[i_XF, F.\widetilde{Y}]_{Cl}.\psi\nonumber\\
&&+k_2^2[i_XF, i_YF]_{Cl}.\psi-k_1F.\widetilde{[X,Y]}.\psi-k_2(i_{[X,Y]}F).\psi.
\end{eqnarray}
In normal coordinates, we have
\begin{eqnarray}
\widehat{R}(X_a,X_b)\psi&=&R(X_a,X_b)\psi+k_1(\nabla_{X_a}F.e_b-\nabla_{X_b}F.e_a).\psi+k_2(\nabla_{X_a}i_{X_b}F-\nabla_{X_b}i_{X_a}F).\psi\nonumber\\
&&+k_1^2F.(e_a.F.e_b-e_b.F.e_a).\psi+k_1k_2\left([F.e_a, i_{X_b}F]_{Cl}+[i_{X_a}F, F.e_b]_{Cl}\right).\psi\nonumber\\
&&+k_2^2[i_{X_a}F, i_{X_b}F]_{Cl}.\psi.
\end{eqnarray}
If we use the following Clifford algebra identities
\begin{eqnarray}
e_a.F.e_b-e_b.F.e_a &=& 2(e_{ab}\wedge F-i_{X_a}i_{X_b}F)\\
{[F.e_a, i_{X_b}F]_{Cl}} &=& 2(e_a\wedge F\wedge i_{X_b}F-i_{X_a}F\wedge i_{X_b}F)\\
{[i_{X_a}F, F.e_b]_{Cl}} &=& 2(i_{X_a}F\wedge e_b\wedge F-i_{X_a}F\wedge i_{X_b}F)\\
{[i_{X_a}F, i_{X_b}F]_{Cl}} &=& 2i_{X_a}F\wedge i_{X_b}F
\end{eqnarray}
we can write (A28) in terms of wedge products as
\begin{eqnarray}
\widehat{R}(X_a,X_b)\psi&=&R(X_a,X_b)\psi+k_1(\nabla_{X_a}F.e_b-\nabla_{X_b}F.e_a).\psi+k_2(\nabla_{X_a}i_{X_b}F-\nabla_{X_b}i_{X_a}F).\psi\nonumber\\
&&+2k_1^2F.(e_{ab}\wedge F-i_{X_a}i_{X_b}F).\psi+2(k_2^2-2k_1k_2)(i_{X_a}F\wedge i_{X_b}F).\psi\nonumber\\
&&+4k_1k_2\left(F\wedge (e_a\wedge i_{X_b}F-e_b\wedge i_{X_a}F)\right).\psi.
\end{eqnarray}
For the choice of $k_1=\frac{1}{2\sqrt{3}}$ and $k_2=-\frac{1}{2\sqrt{3}}$, we obtain the action of the spin curvature operator in five dimensional supergravity theory which reads as
\begin{eqnarray}
\widehat{R}(X_a,X_b)\psi&=&R(X_a,X_b)\psi+\frac{1}{2\sqrt{3}}\left(\nabla_{X_a}F.e_b-\nabla_{X_b}F.e_a\right).\psi-\frac{1}{2\sqrt{3}}\left(\nabla_{X_a}i_{X_b}F-\nabla_{X_b}i_{X_a}F\right).\psi\nonumber\\
&&+\frac{1}{6}F.(e_{ab}\wedge F-i_{X_a}i_{X_b}F).\psi-\frac{1}{6}(i_{X_a}F\wedge i_{X_b}F).\psi\nonumber\\
&&-\frac{1}{3}\left(F\wedge (e_a\wedge i_{X_b}F-e_b\wedge i_{X_a}F)\right).\psi.
\end{eqnarray}

\textit{3-form modification}:

By using the 3-form modified spin connection in (A11), we can write the 3-form modified spin curvature operator as
\begin{eqnarray}
\widehat{R}(X,Y)\psi&=&R(X,Y)\psi+k_1(\nabla_XH.\widetilde{Y}-\nabla_YH.\widetilde{X}).\psi+k_1H.(\nabla_X\widetilde{Y}-\nabla_Y\widetilde{X}).\psi\nonumber\\
&&+k_2(\nabla_Xi_YH-\nabla_Yi_XH).\psi+k_1^2H.(\widetilde{X}.H.\widetilde{Y}-\widetilde{Y}.H.\widetilde{X}).\psi+k_1k_2[H.\widetilde{X}, i_YH]_{Cl}.\psi\nonumber\\
&&k_1k_2[i_XH, H.\widetilde{Y}]_{Cl}.\psi+k_2^2[i_XH, i_YH]_{Cl}.\psi-k_1H.\widetilde{[X, Y]}.\psi-k_2i_{[X,Y]}H.\psi.
\end{eqnarray}
In normal coordinates, we have
\begin{eqnarray}
\widehat{R}(X_a, X_b)\psi&=&R(X_a, X_b)\psi+k_1(\nabla_{X_a}H.e_b-\nabla_{X_b}H.e_a).\psi+k_2(\nabla_{X_a}i_{X_b}H-\nabla_{X_b}i_{X_a}H).\psi\nonumber\\
&&+k_1^2H.(e_a.H.e_b-e_b.H.e_a).\psi+k_1k_2[H.e_a, i_{X_b}H]_{Cl}.\psi+k_1k_2[i_{X_a}H, H.e_b]_{Cl}.\psi\nonumber\\
&&+k_2^2[i_{X_a}H, i_{X_b}H]_{Cl}.\psi
\end{eqnarray}
and by using the following Clifford algebra identities
\begin{eqnarray}
{[H.e_a, i_{X_b}H]_{Cl}}&=&2((e_a\wedge H)\underset{1}{\wedge}i_{X_b}H-i_{X_a}H\underset{1}{\wedge}i_{X_b}H)\\
{[i_{X_a}H, H.e_b]_{Cl}}&=&2(i_{X_a}H\underset{1}{\wedge}(e_b\wedge H)-i_{X_a}H\underset{1}{\wedge}i_{X_b}H)\\
e_a.H.e_b-e_b.H.e_a&=&2(i_{X_a}i_{X_b}H-e_a\wedge e_b\wedge H)\\
{[i_{X_a}H, i_{X_b}H]_{Cl}}&=&-2i_{X_a}H\underset{1}{\wedge}i_{X_b}H
\end{eqnarray}
we have
\begin{eqnarray}
\widehat{R}(X_a, X_b)\psi&=&R(X_a, X_b)\psi+k_1(\nabla_{X_a}H.e_b-\nabla_{X_b}H.e_a).\psi+k_2(\nabla_{X_a}i_{X_b}H-\nabla_{X_b}i_{X_a}H).\psi\nonumber\\
&&+2k_1^2H.(i_{X_a}i_{X_b}H-e_{ab}\wedge H).\psi+2k_1k_2((e_a\wedge H)\underset{1}{\wedge}i_{X_b}H+i_{X_a}H\underset{1}{\wedge}(e_b\wedge H)).\psi\nonumber\\
&&-(4k_1k_2+2k_2^2)(i_{X_a}H\underset{1}{\wedge}i_{X_b}H).\psi
\end{eqnarray}
where we have defined the contracted wedge product for any forms $\alpha$ and $\beta$ as follows
\begin{equation}
\alpha\underset{k}{\wedge}\beta=i_{X_{a_1}}i_{X_{a_2}}...i_{X_{a_k}}\alpha\wedge i_{X_{a^1}}i_{X_{a^2}}...i_{X_{a^k}}\beta.
\end{equation}
For the choice of $k_1=0$ and $k_2=\frac{1}{4}$, we get the modified spin curvature operator in six or ten dimensional supergravity theories as
\begin{equation}
\widehat{R}(X_a, X_b)\psi=R(X_a, X_b)\psi+\frac{1}{4}(\nabla_{X_a}i_{X_b}H-\nabla_{X_b}i_{X_a}H).\psi-\frac{1}{8}(i_{X_a}H\underset{1}{\wedge}i_{X_b}H).\psi.
\end{equation}

\textit{4-form modification}:

From the 4-form modified spin connection in (A15) and similar calculations as in the previous cases, 4-form modified spin curvature operator can be written in normal coordinates as follows
\begin{eqnarray}
\widehat{R}(X_a, X_b)\psi&=&R(X_a, X_b)\psi+k_1(\nabla_{X_a}G.e_b-\nabla_{X_b}G.e_a).\psi+k_2(\nabla_{X_a}i_{X_b}G-\nabla_{X_b}i_{X_a}G).\psi\nonumber\\
&&+k_1^2G.(e_a.G.e_b-e_b.G.e_a).\psi+k_1k_2[G.e_a, i_{X_b}G]_{Cl}.\psi\nonumber\\
&&+k_1k_2[i_{X_a}G, G.e_b]_{Cl}.\psi+k_2^2[i_{X_a}G, i_{X_b}G]_{Cl}.\psi.
\end{eqnarray}
By using the following Clifford algebra identities
\begin{eqnarray}
{[G.e_a, i_{X_b}G]_{Cl}}&=&2(e_a\wedge G-i_{X_a}G)\wedge i_{X_b}G-(e_a\wedge G-i_{X_a}G)\underset{2}{\wedge}i_{X_b}G\\
{[i_{X_a}G, G.e_b]_{Cl}}&=&2i_{X_a}G\wedge (e_b\wedge G-i_{X_b}G)-i_{X_a}G\underset{2}{\wedge}(e_b\wedge G-i_{X_b}G)\\
{[i_{X_a}G, i_{X_b}G]_{Cl}}&=&2i_{X_a}G\wedge i_{X_b}G-i_{X_a}G\underset{2}{\wedge}i_{X_b}G\\
e_a.G.e_b-e_b.G.e_a&=&2(e_a\wedge e_b\wedge G-i_{X_a}i_{X_b}G)
\end{eqnarray}
we obtain
\begin{eqnarray}
\widehat{R}(X_a, X_b)\psi&=&R(X_a, X_b)\psi+k_1(\nabla_{X_a}G.e_b-\nabla_{X_b}G.e_a).\psi+k_2(\nabla_{X_a}i_{X_b}G-\nabla_{X_b}i_{X_a}G).\psi\nonumber\\
&&+2k_1^2G.(e_{ab}\wedge G-i_{X_a}i_{X_b}G).\psi\nonumber\\
&&+2k_1k_2(e_a\wedge G\wedge i_{X_b}G-(e_a\wedge G)\underset{2}{\wedge}i_{X_b}G+i_{X_a}G\wedge e_b\wedge G-i_{X_a}G\underset{2}{\wedge}(e_b\wedge G)).\psi\nonumber\\
&&+2(k_2^2-2k_1k_2)(i_{X_a}G\wedge i_{X_b}G-i_{X_a}G\underset{2}{\wedge}i_{X_b}G).\psi.
\end{eqnarray}
By choosing $k_1=-\frac{1}{12}$ and $k_2=\frac{1}{12}$, we obtain the modified spin curvature operator in eleven dimensional supergravity theory which is written as
\begin{eqnarray}
\widehat{R}(X_a, X_b)\psi&=&R(X_a, X_b)\psi-\frac{1}{12}(\nabla_{X_a}G.e_b-\nabla_{X_b}G.e_a).\psi+\frac{1}{12}(\nabla_{X_a}i_{X_b}G-\nabla_{X_b}i_{X_a}G).\psi\nonumber\\
&&+\frac{1}{72}G.(e_{ab}\wedge G-i_{X_a}i_{X_b}G).\psi\nonumber\\
&&-\frac{1}{72}(e_a\wedge G\wedge i_{X_b}G-(e_a\wedge G)\underset{2}{\wedge}i_{X_b}G+i_{X_a}G\wedge e_b\wedge G-i_{X_a}G\underset{2}{\wedge}(e_b\wedge G)).\psi\nonumber\\
&&+\frac{1}{24}(i_{X_a}G\wedge i_{X_b}G-i_{X_a}G\underset{2}{\wedge}i_{X_b}G).\psi.
\end{eqnarray}

\subsection{Squares of modified Dirac operators}

For the 0-form modified Dirac operator written in (A2), we have
\begin{eqnarray}
\widehat{\displaystyle{\not}D}^2\psi&=&\widehat{\nabla}^2\psi+\frac{1}{4}e^{ab}.R_{ab}.\psi+k^2f^2e^{ab}.e_{ab}.\psi+\frac{k}{2}e^{ab}.((i_{X_a}df)e_b-(i_{X_b}df)e_a).\psi
\end{eqnarray}
where we have used (A21) and (5). From the following Clifford algebra identities
\begin{eqnarray}
e^{ab}.R_{ab}&=&-\mathcal{R}\\
e^{ab}.e_{ab}&=&-n(n-2)\\
e^{ab}.e_b&=&(n-1)e^a\\
e^{ab}.e_a&=&-(n-1)e^b
\end{eqnarray}
we can write
\begin{equation}
\widehat{\displaystyle{\not}D}^2\psi=\widehat{\nabla}^2\psi-\frac{1}{4}\mathcal{R}\psi+(n-1)kdf.\psi-n(n-2)k^2f^2\psi.
\end{equation}

For the 1-form modified case, we have from (A25)
\begin{eqnarray}
\widehat{\displaystyle{\not}D}^2\psi&=&\widehat{\nabla}^2\psi+\frac{1}{4}e^{ab}.R_{ab}.\psi-\frac{k_2}{2}e^{ab}(i_{X_a}i_{X_b}dA).\psi+\frac{k_1}{2}e^{ab}.(\nabla_{X_a}A.e_b-\nabla_{X_b}.e_a).\psi\nonumber\\
&&+\frac{k_1^2}{2}e^{ab}.A.(e_a.A.e_b-e_b.A.e_a).\psi.
\end{eqnarray}
If we use the following Clifford algebra identities
\begin{eqnarray}
e^{ab}.\nabla_{X_a}A.e_b&=&-(n-3)dA+(n-1)\delta A\\
e^{ab}.\nabla_{X_b}A.e_a&=&(n-4)dA-n\delta A\\
e^{ab}.A.e_a.A.e_b-e^{ab}.A.e_b.A.e_a&=&(2n-6)(e^a\wedge A).A.e_a+(2n-2)i_{X^a}A.A.e_a
\end{eqnarray}
we can write (A57) as
\begin{eqnarray}
\widehat{\displaystyle{\not}D}^2\psi&=&\widehat{\nabla}^2\psi-\frac{1}{4}\mathcal{R}\psi+(k_2-(n-3)k_1)dA.\psi-(n-1)k_1\delta A.\psi\nonumber\\
&&+(n(n-3)-2(n-2))k_1^2(A\wedge A).\psi+(n(n-3)+2)k_1^2(A\underset{1}{\wedge}A).\psi.
\end{eqnarray}
and by choosing $k_1=0$ and $k_2=1$, we get the square of the gauged Dirac operator
\begin{equation}
\widehat{\displaystyle{\not}D}^2\psi=\widehat{\nabla}^2\psi-\frac{1}{4}\mathcal{R}\psi+dA.\psi.
\end{equation}

The square of the 2-form modified Dirac operator is written from (A28) as
\begin{eqnarray}
\widehat{\displaystyle{\not}D}^2\psi&=&\widehat{\nabla}^2\psi+\frac{1}{4}e^{ab}.R_{ab}.\psi+\frac{k_1}{2}e^{ab}.(\nabla_{X_a}F.e_b-\nabla_{X_b}F.e_a).\psi+\frac{k_2}{2}e^{ab}.(\nabla_{X_a}i_{X_b}F-\nabla_{X_b}i_{X_a}F).\psi\nonumber\\
&&+\frac{k_1^2}{2}e^{ab}.F.(e_a.F.e_b-e_b.F.e_a).\psi+\frac{k_1k_2}{2}e^{ab}.([F.e_a,i_{X_b}F]_{Cl}+[i_{X_a}F,F.e_b]_{Cl}).\psi\nonumber\\
&&+\frac{k_2^2}{2}e^{ab}.[i_{X_a}F,i_{X_b}F]_{Cl}.\psi
\end{eqnarray}
and from the following Clifford algebra identities
\begin{eqnarray}
e^{ab}.(\nabla_{X_a}F.e_b-\nabla_{X_b}F.e_a)&=&2(n-5)dF-2(n-3)\delta F\\
e^{ab}.(\nabla_{X_a}i_{X_b}F-\nabla_{X_b}i_{X_a}F)&=&4dF-2\delta F\\
e^{ab}.F.(e_a.F.e_b-e_b.F.e_a)&=&-2(n-4)(n-9)F\wedge F+2(n-4)(n-5)F\underset{1}{\wedge}F\nonumber\\
&&+(n-1)(n-4)F\underset{2}{\wedge}F\\
e^{ab}.([F.e_a,i_{X_b}F]_{Cl}+[i_{X_a}F,F.e_b]_{Cl})&=&-8(n-6)F\wedge F-4F\underset{1}{\wedge}F+4F\underset{2}{\wedge}F\\
e^{ab}.[i_{X_a}F,i_{X_b}F]_{Cl}&=&-8F\wedge F-6F\underset{1}{\wedge}F-2F\underset{2}{\wedge}F
\end{eqnarray}
we can write
\begin{eqnarray}
\widehat{\displaystyle{\not}D}^2\psi&=&\widehat{\nabla}^2\psi-\frac{1}{4}\mathcal{R}\psi+((n-5)k_1+2k_2)dF.\psi-((n-3)k_1+k_2)\delta F.\psi\nonumber\\
&&-\bigg((n-4)(n-9)k_1^2+4(n-6)k_1k_2+4k_2^2\bigg)(F\wedge F).\psi\nonumber\\
&&+\bigg((n-4)(n-5)k_1^2-2k_1k_2-3k_2^2\bigg)(F\underset{1}{\wedge}F).\psi\nonumber\\
&&+\bigg(\frac{(n-1)(n-4)}{2}k_1^2+2k_1k_2-k_2^2\bigg)(F\underset{2}{\wedge}F).\psi.
\end{eqnarray}
For the choice of $k_1=\frac{1}{2\sqrt{3}}$ and $k_2=-\frac{1}{2\sqrt{3}}$ in five dimensions, we get the square of the 2-form modified Dirac operator in five dimensional supergravity
\begin{eqnarray}
\widehat{\displaystyle{\not}D}^2\psi&=&\widehat{\nabla}^2\psi-\frac{1}{4}\mathcal{R}\psi-\frac{1}{\sqrt{3}}dF.\psi-\frac{1}{2\sqrt{3}}\delta F.\psi\nonumber\\
&&-\frac{1}{3}(F\wedge F).\psi-\frac{1}{12}(F\underset{1}{\wedge}F).\psi-\frac{1}{12}(F\underset{2}{\wedge}F).\psi.
\end{eqnarray}

By considering (A36), we can calculate the square of the 3-form modified Dirac operator as
\begin{eqnarray}
\widehat{\displaystyle{\not}D}^2\psi&=&\widehat{\nabla}^2\psi+\frac{1}{4}e^{ab}.R_{ab}.\psi+\frac{k_1}{2}e^{ab}.(\nabla_{X_a}H.e_b-\nabla_{X_b}H.e_a).\psi+\frac{k_2}{2}e^{ab}.(\nabla_{X_a}i_{X_b}H-\nabla_{X_b}i_{X_a}H).\psi\nonumber\\
&&+\frac{k_1^2}{2}e^{ab}.H.(e_a.H.e_b-e_b.H.e_a).\psi+\frac{k_1k_2}{2}e^{ab}.\left([H.e_a, i_{X_{b}H}]_{Cl}+[i_{X_a}H, H.e_b]_{Cl}\right).\psi\nonumber\\
&&+\frac{k_2^2}{2}e^{ab}.[i_{X_a}H, i_{X_b}H]_{Cl}.\psi.
\end{eqnarray}
If we replace the following Clifford algebra identities in (A71)
\begin{eqnarray}
e^{ab}.(\nabla_{X_a}H.e_b-\nabla_{X_b}H.e_a)&=&-2(n-7)dH+2(n-5)\delta H\\
e^{ab}.(\nabla_{X_a}i_{X_b}H-\nabla_{X_b}i_{X_a}H)&=&6dH+4\delta H\\
e^{ab}.H.(e_a.H.e_b-e_b.H.e_a)&=&2(n-6)(n-13)H\wedge H+2(n-6)(n-9)H\underset{1}{\wedge}H\nonumber\\
&&-(n-6)(n-5)H\underset{2}{\wedge}H-\frac{(n-6)(n-1)}{3}H\underset{3}{\wedge}H\\
e^{ab}.([H.e_a, i_{X_b}H]_{Cl}+[i_{X_a}H, H.e_b]_{Cl})&=&36H\wedge H-8(n-8)H\underset{1}{\wedge}H+2(n-5)H\underset{2}{\wedge}H+4H\underset{3}{\wedge}H\\
e^{ab}.[i_{X_a}H, i_{X_b}H]_{Cl}&=&-18H\wedge H+8H\underset{1}{\wedge}H+H\underset{2}{\wedge}H+2H\underset{3}{\wedge}H
\end{eqnarray}
then we can find
\begin{eqnarray}
\widehat{\displaystyle{\not}D}^2\psi&=&\widehat{\nabla}^2\psi-\frac{1}{4}\mathcal{R}\psi+(3k_2-(n-7)k_1)dH.\psi+((n-5)k_1+2k_2)\delta H.\psi\nonumber\\
&&+\bigg((n-6)(n-13)k_1^2+18k_1k_2-9k_2^2\bigg)(H\wedge H).\psi\nonumber\\
&&+\bigg((n-6)(n-9)k_1^2-4(n-8)k_1k_2+4k_2^2\bigg)(H\underset{1}{\wedge}H).\psi\nonumber\\
&&+\bigg(-\frac{(n-6)(n-5)}{2}k_1^2+(n-5)k_1k_2+\frac{k_2^2}{2}\bigg)(H\underset{2}{\wedge}H).\psi\nonumber\\
&&+\bigg(-\frac{(n-6)(n-1)}{6}k_1^2+2k_1k_2+k_2^2\bigg)(H\underset{3}{\wedge}H).\psi.
\end{eqnarray}
By choosing $k_1=0$ and $k_2=\frac{1}{4}$, we get the square of the 3-form modified Dirac operator in six or ten dimensional supergravity theories
\begin{eqnarray}
\widehat{\displaystyle{\not}D}^2\psi&=&\widehat{\nabla}^2\psi-\frac{1}{4}\mathcal{R}\psi+\frac{3}{4}dH.\psi+\frac{1}{2}\delta H.\psi-\frac{9}{16}(H\wedge H).\psi\nonumber\\
&&+\frac{1}{4}(H\underset{1}{\wedge}H).\psi+\frac{1}{32}(H\underset{2}{\wedge}H).\psi+\frac{1}{16}(H\underset{3}{\wedge}H).\psi.
\end{eqnarray}

We can also write the square of the 4-form modified Dirac operator for a 4-form $G$ from (A44) as
\begin{eqnarray}
\widehat{\displaystyle{\not}D}^2\psi&=&\widehat{\nabla}^2\psi+\frac{1}{4}e^{ab}.R_{ab}.\psi+\frac{k_1}{2}e^{ab}.(\nabla_{X_a}G.e_b-\nabla_{X_b}G.e_a).\psi+\frac{k_2}{2}e^{ab}.(\nabla_{X_a}i_{X_b}G-\nabla_{X_b}i_{X_a}G).\psi\nonumber\\
&&+\frac{k_1^2}{2}e^{ab}.G.(e_a.G.e_b-e_b.G.e_a).\psi+\frac{k_1k_2}{2}e^{ab}.\left([G.e_a, i_{X_{b}G}]_{Cl}+[i_{X_a}G, G.e_b]_{Cl}\right).\psi\nonumber\\
&&+\frac{k_2^2}{2}e^{ab}.[i_{X_a}G, i_{X_b}G]_{Cl}.\psi
\end{eqnarray}
and by considering the following Clifford algebra identities
\begin{eqnarray}
e^{ab}.(\nabla_{X_a}G.e_b-\nabla_{X_b}G.e_a)&=&2(n-9)dG-2(n-7)\delta G\\
e^{ab}.(\nabla_{X_a}i_{X_b}G-\nabla_{X_b}i_{X_a}G)&=&8dG-6\delta G\\
e^{ab}.G.(e_a.G.e_b-e_b.G.e_a)&=&-2(n-8)(n-15)G\wedge G+2(n-8)(n-11)G\underset{1}{\wedge}G\nonumber\\
&&+(n-8)(n-7)G\underset{2}{\wedge}G-\frac{(n-8)(n-5)}{3}G\underset{3}{\wedge}G\nonumber\\
&&-\frac{(n-8)(n-1)}{12}G\underset{4}{\wedge}G\\
e^{ab}.([G.e_a, i_{X_b}G]_{Cl}+[i_{X_a}G, G.e_b]_{Cl})&=&-16(n-12)G\wedge G-36G\underset{1}{\wedge}G+4(n-9)G\underset{2}{\wedge}G\nonumber\\
&&-\frac{16}{3}G\underset{3}{\wedge}G-2G\underset{4}{\wedge}G\\
e^{ab}.[i_{X_a}G, i_{X_b}G]_{Cl}&=&-32G\wedge G-30G\underset{1}{\wedge}G+2G\underset{2}{\wedge}G+\frac{13}{3}G\underset{3}{\wedge}G+G\underset{4}{\wedge}G
\end{eqnarray}
we can find
\begin{eqnarray}
\widehat{\displaystyle{\not}D}^2\psi&=&\widehat{\nabla}^2\psi-\frac{1}{4}\mathcal{R}\psi+((n-9)k_1+4k_2)dG.\psi-((n-7)k_1+3k_2)\delta G.\psi\nonumber\\
&&+\bigg(-(n-8)(n-15)k_1^2-8(n-12)k_1k_2-16k_2^2\bigg)(G\wedge G).\psi\nonumber\\
&&+\bigg((n-8)(n-11)k_1^2-18k_1k_2-15k_2^2\bigg)(G\underset{1}{\wedge}G).\psi\nonumber\\
&&+\bigg(\frac{(n-8)(n-7)}{2}k_1^2+2(n-9)k_1k_2+k_2^2\bigg)(G\underset{2}{\wedge}G).\psi\nonumber\\
&&+\bigg(-\frac{(n-8)(n-5)}{6}k_1^2-\frac{8}{3}k_1k_2+\frac{13}{6}k_2^2\bigg)(G\underset{3}{\wedge}G).\psi\nonumber\\
&&+\bigg(-\frac{(n-8)(n-1)}{24}k_1^2-k_1k_2+\frac{k_2^2}{2}\bigg)(G\underset{4}{\wedge}G).\psi.
\end{eqnarray}
If we choose $k_1=-\frac{1}{12}$ and $k_2=\frac{1}{12}$ for $n=11$, we get the square of the modified Dirac operator in eleven dimensional supergravity
\begin{eqnarray}
\widehat{\displaystyle{\not}D}^2\psi&=&\widehat{\nabla}^2\psi-\frac{1}{4}\mathcal{R}\psi+\frac{1}{6}dG.\psi+\frac{1}{12}\delta G.\psi-\frac{1}{12}(G\wedge G).\psi+\frac{1}{48}(G\underset{1}{\wedge}G).\psi\nonumber\\
&&+\frac{1}{48}(G\underset{2}{\wedge}G).\psi+\frac{11}{864}(G\underset{3}{\wedge}G).\psi+\frac{1}{576}(G\underset{4}{\wedge}G).\psi.
\end{eqnarray}\\

\end{appendix}


\end{document}